\documentclass{nature_custom}
\usepackage{graphicx,amssymb}
\usepackage{amsmath}
\usepackage{mathdots}
\usepackage{txfonts}
\usepackage[classicReIm]{kpfonts}
\usepackage{lineno,hyperref}
\usepackage{siunitx}
\usepackage{braket}
\modulolinenumbers[1]

\usepackage{siunitx}

\newcommand{\SNR}{\text{SNR}}
\newcommand{\SNRc}{\text{SNR}_\text{0}}

\newcommand{\Ns}{N_\text{S}}
\newcommand{\Nb}{N_\text{B}}

\newcommand{\Deff}{D_\text{eff}}
\newcommand{\Deffc}{D_\text{eff,0}}
\newcommand{\Ts}{T_\text{S}}
\newcommand{\Tb}{T_\text{B}}

\newcommand{\Ninel}{N_\text{inel}}
\newcommand{\tinel}{t_\text{inel}}

\newcommand{\HIV}{\text{HIV-1 Gag}}
\newcommand{\Mar}{\text{MARV VP35}}

\newcommand{\figref}[1]{Fig. \ref{#1}}
\newcommand{\figsref}[1]{Figs. \ref{#1}}

\usepackage{affils}

\title{Multi-pass transmission electron microscopy}

\author{Thomas Juffmann$^{1}$, Stewart A. Koppell$^1$, Brannon B. Klopfer $^1$, Colin Ophus$^2$, Robert Glaeser$^3$ \& Mark A. Kasevich$^1$}
\begin{document}
\maketitle
\begin{affiliations}
 \item Physics Department, Stanford University, 382 Via Pueblo Mall, Stanford, California 94305, USA
 \item National Center for Electron Microscopy, Molecular Foundry, Lawrence Berkeley National Laboratory, 1 Cyclotron Road, Berkeley, California 94720, USA
 \item Molecular Biophysics and Integrative Bioimaging, Lawrence Berkeley National Laboratory, University of California, Berkeley, California 94720, USA
\end{affiliations}

\begin{abstract}
Feynman once asked physicists to build better electron microscopes to be able to watch biology at work. While electron microscopes can now provide atomic resolution, electron beam induced specimen damage precludes high resolution imaging of sensitive materials, such as single proteins or polymers. Here, we use simulations to show that an electron microscope based on a simple multi-pass measurement protocol enables imaging of single proteins at reduced damage and at nanometer resolution, without averaging structures over multiple images. While we demonstrate the method for particular imaging targets, the approach is broadly applicable and is expected to improve resolution and sensitivity for a range of electron microscopy imaging modalities, including, for example, scanning and spectroscopic techniques. The approach implements a quantum mechanically optimal strategy which under idealized conditions can be considered interaction-free.  In practice, an order-of-magnitude reduction in damage at equivalent resolution appears feasible.

\end{abstract}

%\linenumbers

\clearpage

Only a finite number of electrons can be used to probe a biological specimen before damaging the structure of interest \cite{Egerton2004}. In conjunction with electron counting statistics (shot-noise), this leads to a finite signal-to-noise ratio (SNR) and a spatial resolution which is not limited by the quality of the electron optics, but rather by the sample-specific maximally allowed electron dose. For typical proteins imaged using cryo electron microscopy the achievable spatial resolution is about \SI{2}{nm} assuming ideal instrumentation \cite{Glaeser2011ReachingAspects}. To reconstruct a protein model at atomic resolution, thousands of images of single proteins have to be averaged \cite{Henderson1995, Glaeser2016a}. This process is time consuming and potentially erroneous \cite{Henderson2013AvoidingNoise.} and usually assumes that all imaged proteins are structurally identical. For polymers, heterogeneous organic molecules and other forms of aperiodic beam-sensitive soft matter, averaging techniques are not applicable. 
Previous research has focused on approaching the dose limited resolution (DLR) is given by shot-noise and the critical dose of a specimen. Significant progress has recently been achieved \cite{Nogales2016,Fernandez-Leiro2016UnravellingMicroscopy} with novel detectors \cite{Bai2015}, electron optical elements \cite{Danev2014,Khoshouei2016VoltaPrx3}, and sample preparation techniques \cite{Chamberlain2015Transmission5/2015,Ross2015OpportunitiesMicroscopy}, and the optimum classical DLR may soon be reached \cite{Glaeser2016a}. Conceptually new forms of microscopy will be required to push electron microscopy further.

In transmission electron microscopy (TEM), biological specimens manifest as weak phase objects. The achievable sensitivity in phase measurements has been discussed in the quantum measurement community\cite{Giovannetti2004a}.  Using uncorrelated probe particles, the lowest achievable measurement error is $1/\sqrt[]{N}$, where $N$ is the number of probe particle-sample interactions. This so-called shot-noise limit can be overcome using correlated particles, and the error can be reduced to $1/N$, the Heisenberg limit\cite{Giovannetti2011a}. Adequately entangled photons provide these correlations and have been applied in optical microscopes\cite{Ono2013a,Israel2014a}. However these entangled states are difficult to create and the most commonly discussed N00N states \cite{Dowling2008} rely on the bosonic nature of photons. While one can conceive entangled (hybrid) systems that allow approaching the Heisenberg limit with fermions \cite{Yurke1986InputSensitivity,Okamoto2014}, these appear difficult to implement experimentally. However, it can also be approached with a single probe particle which interacts with the phase object multiple times \cite{Higgins2007a} and it was shown that this is an optimal measurement strategy at a given number of probe particle-sample interactions \cite{Giovannetti2006a}. Using self-imaging cavities \cite{Arnaud1969a} this approach has recently been extended to full field optical microscopy \cite{Juffmann2016Multi-passMicroscopy,Klopfer2016IterativeLight}. 

In this paper we demonstrate through simulations that a multi-pass protocol can enhance the sensitivity and spatial resolution of dose-limited TEM. Multi-pass TEM image simulations of protein structures embedded in vitreous ice demonstrate order-of-magnitude improvements in typical cryo-EM experiments, and simulations of single-layer graphene images illustrate the limits of the multi-pass technique. 

\section{Results}
\subsection{Reduced damage using multi-pass microscopy.}
\begin{figure}
	\centering
	 \includegraphics[width=8.8cm]{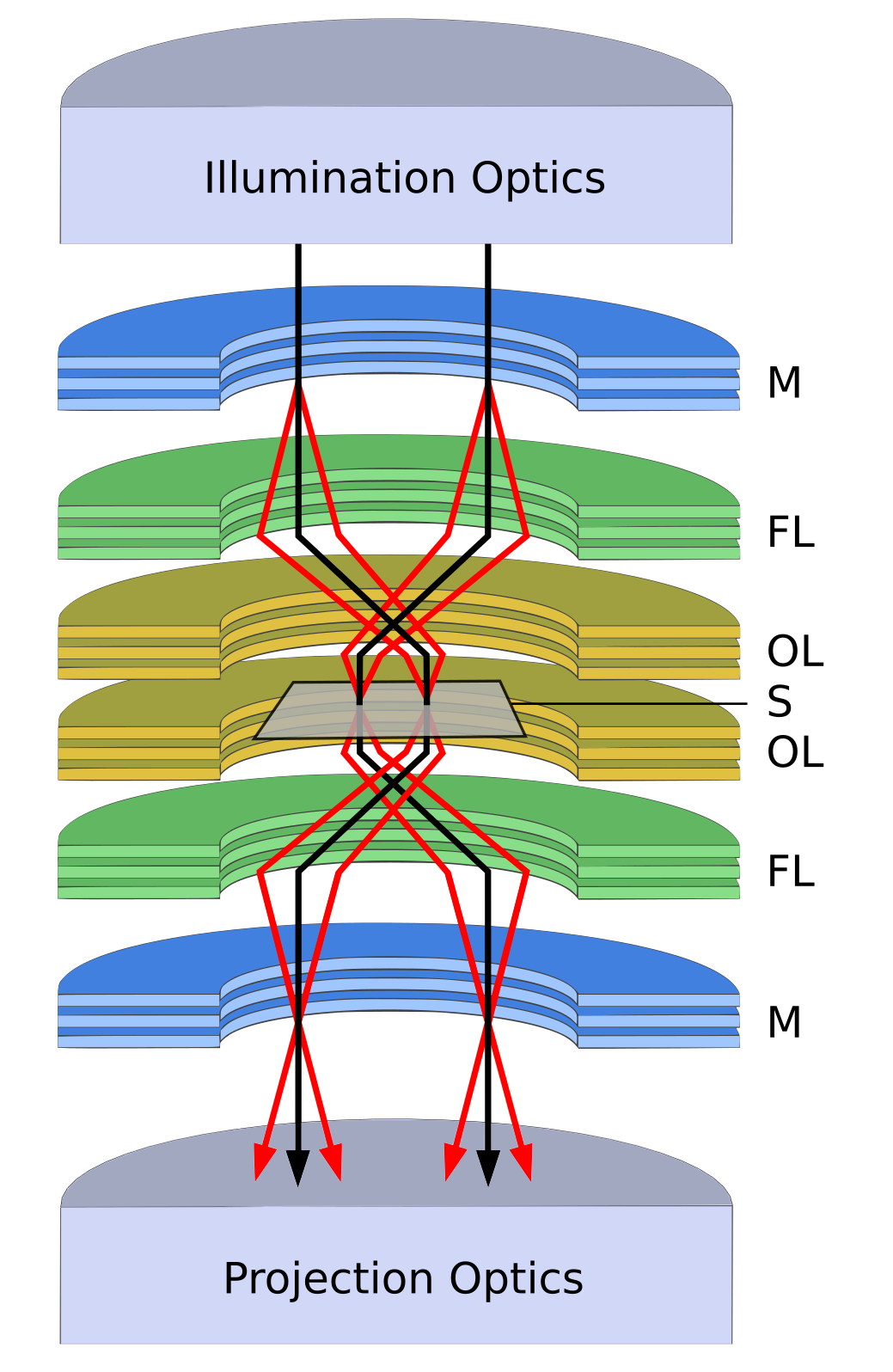}
	\caption{\textbf{Schematic of multi-pass microscopy.} A sample S is placed between two two objective and field lenses (OL and FL, respectively). This configuration is placed in between two mirrors M, which can be gated for in- and out-coupling of the electron beam (see methods). A pulsed probe beam is coupled into the optical path of the multi-pass microscope and illuminates S. The exit wave is subsequently re-imaged back onto the sample, which is now illuminated with an in-focus image of itself. This process is repeated multiple (\textit{m}) times, after which the pulse is out-coupled and imaged onto a detector. For illustration, field (black) and imaging (red) rays are shown, which retrace themselves after one full roundtrip.  \label{fig:s1}}
\end{figure}

A sketch of a multi-pass TEM is shown in \figref{fig:s1}. The image formed by an aberration-free implementation
can be obtained through iterative application of the single pass transmission function $t$ of the sample.  For $m$ passes, the effective transmission function $t_m$ becomes $t_m=t^m=|t|^me^{im\phi}$, where $|t|$ is the transmission magnitude and $\phi$ is the phase shift induced by its potential, both of which vary spatially.
In a phase microscope \cite{Zernike1942,Danev2014} a highly transmissive  ($1-|t|^m\ll 1$) and weak ($m\phi\ll 1$) specimen will yield $N(x,y)\sim N_0\left[1-2m\phi\left(x,y\right)\right]$  detected electrons \cite{Born1993}, with $N_0$ electrons illuminating an area $\delta^2$ that is imaged onto a single pixel of the detector. A multi-pass configuration thus leads to an $m$ fold signal and sensitivity enhancement, while shot noise is $\sim \sqrt[]{N(x,y)}$.
The signal to noise ratio becomes $\SNR=\left|\Ns-\Nb\right|/\sqrt{\Ns+\Nb} \sim \sqrt[]{2N_0}m\Delta\phi$, where  $\Ns$ and $\Nb$ give the number of detected electrons when imaging the specimen and background, respectively. 

For operation at constant damage the number of incoming probe particles has to be chosen such that the total number of probe-particle sample interactions is independent of $m$ -- e.g. $N_{0,m}\sim N_{0,1}/m$. This yields a $\SNR$ at constant damage proportional to $\sqrt[]{m}$ and, alternately, a damage reduction at constant $\SNR$ proportional to $1/m$. 
This also holds for scattering contrast and dark-field detection techniques (see methods).

Under idealized conditions, the multi-pass method has similar damage scaling as  interaction-free methods \cite{Elitzur1993,Kwiat1995,Putnam2009,Kruit2016} previously proposed for the non-destructive detection of fully absorbing samples.  To demonstrate this, we consider the threshold $\SNR$ for detection of a phase object to be $\SNR=\sqrt[]{2N_0}m\Delta\phi \sim 1$. On the other hand, the number of electrons that cause damage by scattering inelastically is $\Ninel=N_0\left(1-|\tinel|^{2m}\right)\sim 2N_0m\alpha$, where $\alpha=1-|\tinel|$, and elastic losses are assumed to be negligible (i.e. no electrons are scattered out of the aperture of the microscope). The quantum interaction-free regime is reached for $\Ninel=\alpha/m\Delta\phi^2\ll1$.  Thus the multi-pass configuration can be considered as the interaction-free configuration appropriate for 
weak specimens. While original interaction-free protocols \cite{Elitzur1993,Kwiat1995} were designed for detecting perfect absorbers, the multi-pass method is capable of providing both phase and gray-scale (see methods) information on highly transparent samples. As we demonstrate below, multi-pass TEM allows for significant damage reduction in the imaging of thin, biological specimens under realistic conditions.
However, we emphasize that truly interaction-free imaging should be considered as a theoretical limit only.

Reduced damage directly translates into improved dose limited spatial resolution (DLR).  
Since the $\SNR$ at constant damage is proportional to $\sqrt{N_0m}\Delta\phi$, this suggests that even at $m=1$ the smallest phase objects could be detected with high $\SNR$ as long as $N_0$ is large enough. However, as radiation  can destroy the structural features of interest, images are often acquired at a single-pass dose $ D=\frac{eN_0}{{\delta }^2}$ about twice the critical dose $D_c$ \cite{Hayward1979RadiationTemperature,Egerton2014a}.
 This leads to a minimum feature size $\delta$ that can be imaged with a given $\SNR$. Using the above equations we see that $\delta \propto 1/\sqrt[]{m}$. This proportionality also holds for scattering contrast (see methods).

\subsection{Multi-pass TEM simulations.}
In the following we show multi-pass TEM simulations  of three model systems of known structure: graphene \cite{Boehm1962DasKohlenstoff-Folien,Novoselov2004ElectricFilms}, the hexameric unit of the immature HIV-1 Gag CTD-SP1 lattice ($\HIV$, PDB ID: 5I4T) \cite{Wagner2016CrystalSwitch.}, and the Marburg Virus VP35 Oligomerization Domain P4222 ($\Mar$, PDB ID: 5TOI) \cite{Bruhn2016CrystalDomain.}.
In the simulations (see methods for details) an electron wave passes through the sample of an aberration-free multi-pass TEM multiple times. After $m$ passes, the resulting exit wave is imaged onto an ideal detector. We consider a phase sensitive detection scheme employing a phase plate to shift the phase of the undiffracted beam by $\pm\pi/2$.  This can be realized with various techniques \cite{Zernike1942,Boersch1947,Schwartz2016ContinuousCavity_specialformatting,Glaeser2013a,Danev2014}.  Poissonian noise is applied to the detected intensity to simulate shot-noise.
The incoming electron dose is chosen such that the effective dose, i.e. the number of electron-sample interactions and thus the electron induced damage, is independent of $m$. For a lossless sample this implies that the incoming dose is scaled by $1/m$. In the simulations, both elastic and inelastic loss is considered (see methods). 
 
\begin{figure}
\centering
\includegraphics[width=8.8cm]{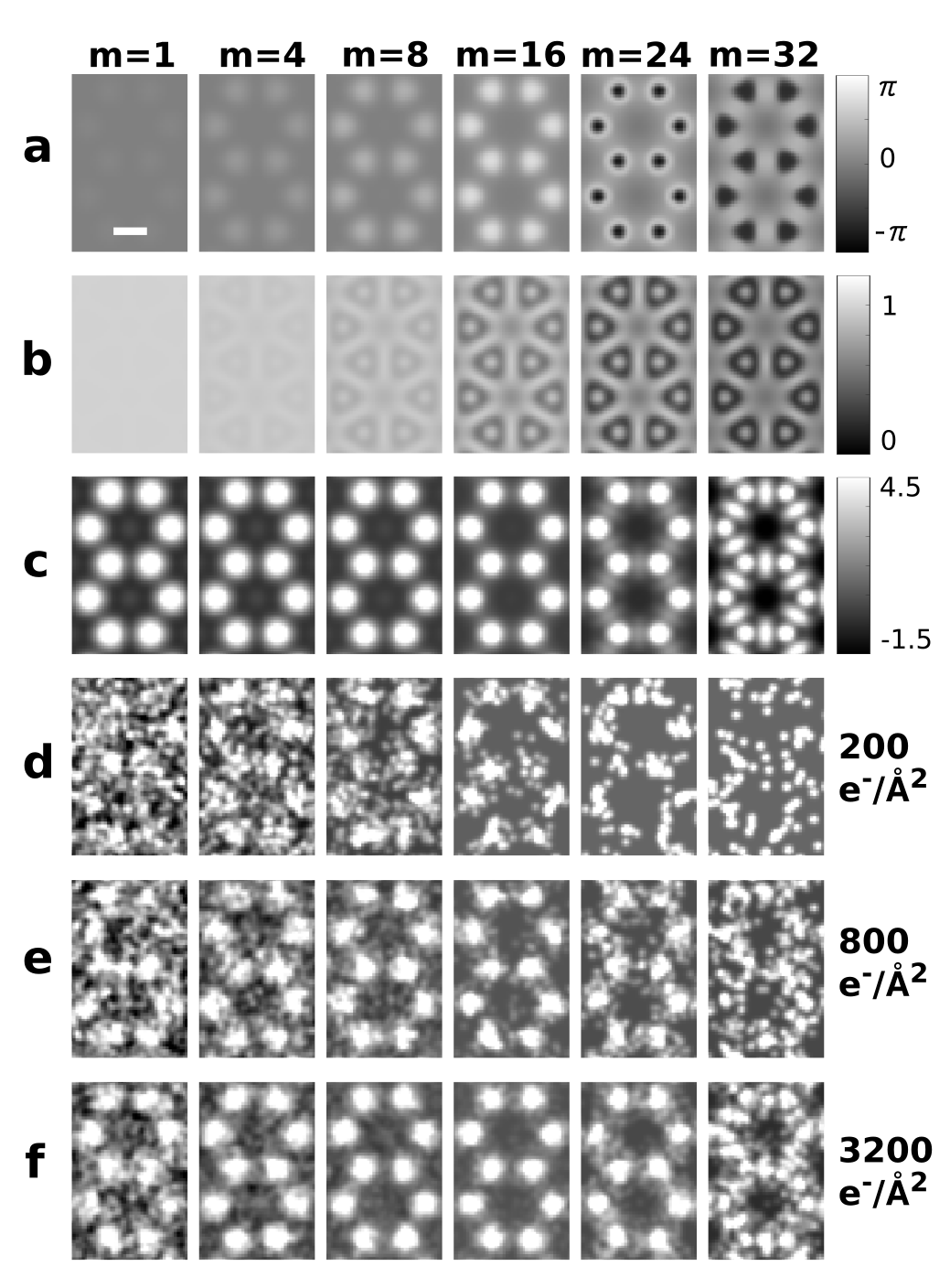}
	\caption{\textbf{Multi-pass TEM simulation of graphene.} (a) and (b) show the phase and amplitude of the exit wave function after a given number of passes, respectively. Simulated multi-pass phase TEM images, noise-free (c), and at various effective dose levels (d-f). 
The colorscale for (c-f) is in units of standard deviations from the mean intensity of each image. The scalebar is \SI{0.14}{nm}.} \label{fig:graphene} 
\end{figure}
The simulations for graphene were done with an electron energy of \SI{60}{keV}, chosen to be low enough to minimize damage \cite{Meyer2012AccurateGraphene}. 
\figsref{fig:graphene} (a) and (b) show the phase and amplitude (respectively) of the simulated exit wave function as a function of the number of interactions. The phase shifts build up linearly, eventually to more than $\pi$. The amplitude of the exit wave function decreases with the number of interactions. Although loss is assumed to be homogeneous across the unit cell\cite{Lee2013TheSilicon}, the lattice structure becomes apparent at higher interaction numbers. This is because the spatially distributed phase shifts cause significant lensing. In this regime, phase contrast is transferred into amplitude contrast even in absence of a phase plate. A noise-free image of the exit wave function is shown in \figref{fig:graphene} (c). The detrimental effect of counting statistics on spatial resolution becomes apparent in \figsref{fig:graphene} (d-f), which show simulated images as a function of effective dose. While in \figsref{fig:graphene} (d) and (e) the lattice structure is not visible after a single interaction, multiple passes improve the $\SNR$ and therefore the spatial resolution. At higher interaction numbers the $\SNR$ decreases again, mainly because phase shifts build up to an extent that standard phase microscopy is no longer the ideal read-out scheme, an effect that also becomes apparent in \figref{fig:graphene} (c). Electron losses also reduce the visibility at higher interaction numbers.  

\begin{figure}
\centering
\includegraphics[width=18cm]{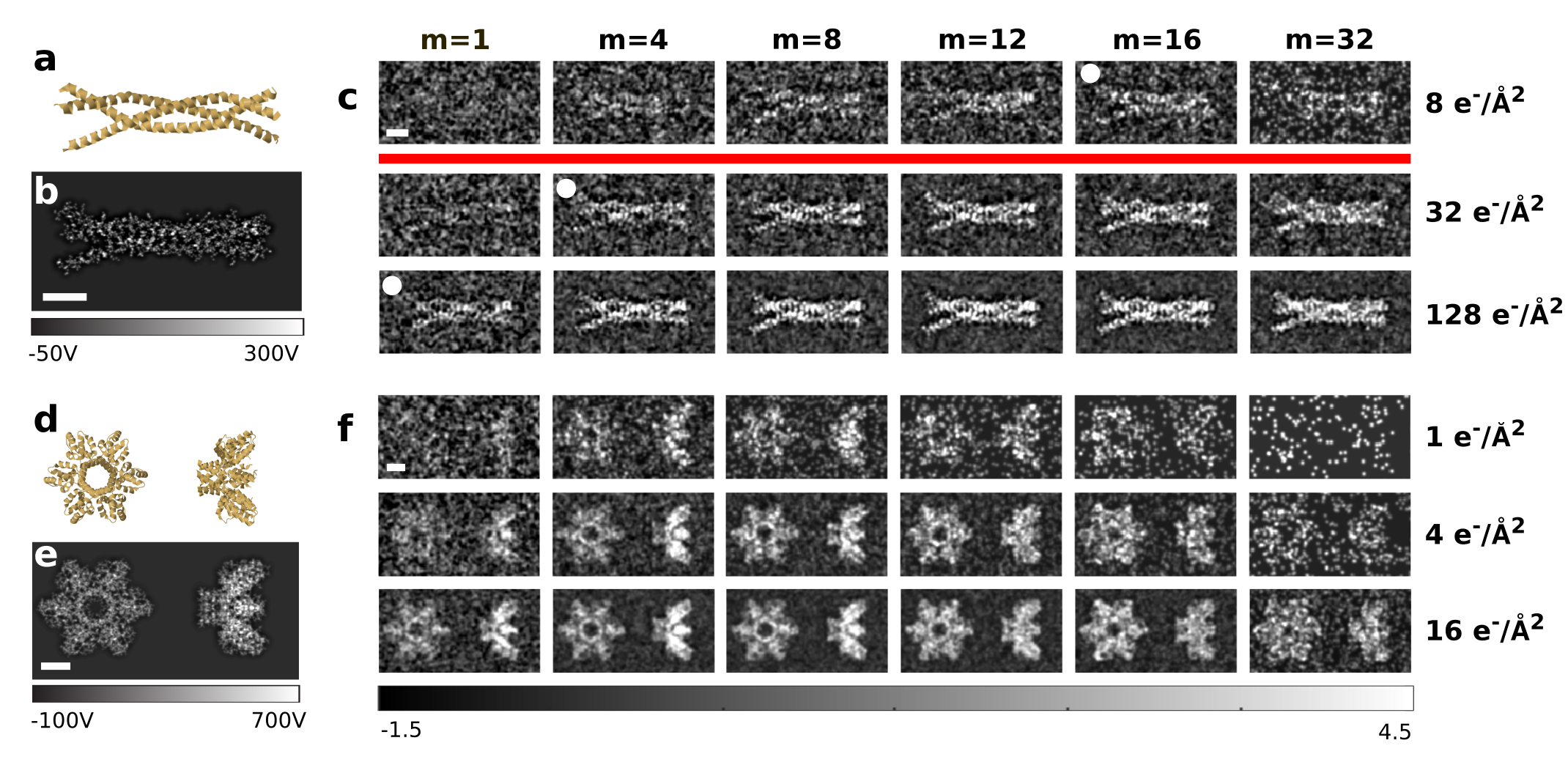}
	\caption{\textbf{Multi-pass TEM simulation of protein structures.} (a) shows the ribbon diagram and (b) shows the projected potential of $\Mar$. (c) shows simulated multi-pass phase TEM images for 300 keV electrons, calculated at three respective levels of effective dose, which are shown on the right. Note that the incoming dose for each panel is roughly m-fold lower than the effective dose, as is explained in the text. 
The white circles indicate figures of similar $\SNR$ (see text). The red line indicates the typical critical dose ($\sim \SI{20}{e^{-}/\angstrom^2}$) for biological specimens \cite{Egerton2014a}. (d-f) show the results for two different projections of $\HIV$.  All scalebars are \SI{2}{nm}, the colorscale for (c) and (f) is in units of standard deviations from the mean intensity of each image. \label{fig:protein}}
\end{figure}

\figsref{fig:protein} (a) and (b) show the ribbon diagram and projected potential of \Mar, which is embedded in \SI{20}{nm} of vitreous ice for cryo electron microscopy. Inelastic losses are dominated by scattering in the vitreous ice, which has an inelastic mean free path of \SI{350}{nm} for electrons at \SI{300}{keV} \cite{Vulovic2013ImageMicroscopy}.
\figref{fig:protein} (c) shows simulated multi-pass TEM results at various effective dose levels. For a given effective dose ($\Deff$) the image quality improves with the number of passes. The best $\SNR$ is achieved after about 12 to 16 passes, where a single alpha helix becomes apparent at a dose below the critical dose for biological specimens. For a higher number of passes the $\SNR$ decreases again, both due to phase build-up and inelastic losses. \figref{fig:protein} (c) also shows the $1/m$ damage reduction at constant $\SNR$. The image at ($m=1,\Deff=\SI{128}{e^{-}/\angstrom^2}$) has a $\SNR$ equivalent to to the one at ($m=4,\Deff=\SI{32}{e^{-}/\angstrom^2}$), and at ($m=16, \Deff=\SI{8}{e^{-}/\angstrom^2}$) before taking losses into account.

\figsref{fig:protein} (d-f) show simulations for $\HIV$ in two different orientations. Due to phase wrapping, the best $\SNR$ is now achieved after 8 to 12 (4 to 8) passes for the projection along the thin (thick) axis of the protein, respectively. For such medium sized proteins, multi-pass microscopy enables the identification of the protein orientation at extremely low dose. One important application of this might be to record dose-fractionated movies with lower effective exposure levels per frame compared to what is currently needed to align successive frames. The reason to do so is that beam-induced movement is much greater over the first 2 to \SI{4}{e^{-}/\angstrom^2} of an exposure, while, at the same time, the high-resolution features of a specimen are rapidly becoming damaged during that time \cite{Scheres2014Beam-inducedParticles.}.  
Reduction of frame-to-frame motion is expected to retain most of the high-resolution signal that is currently lost due to beam-induced motion.

\section{Discussion}
Our analysis shows that the signal enhancement provided by multi-pass protocols can enable the detection of highly transmissive specimens at minimal damage. We have shown that details of dose sensitive specimens can be revealed without averaging, under realistic imaging conditions. Multi-pass TEM offers a quantum optimal approach to the study of, for example, single proteins, DNA, and polymers. 

\begin{methods}

\subsection{Scattering Contrast (Gray-Scale) Multi-Pass TEM.}
In scattering contrast TEM, contrast is obtained from spatially varying electron loss due to elastic and inelastic scattering events. 
Scattering contrast is insensitive to weak phase shifts. A local and real transmission $T$ of the sample can then be defined based on ${\lambda }_f$, the mean free path length in between scattering events that lead to loss:
\begin{equation}
T\left({\alpha }_0\right)=|t\left({\alpha }_0\right)|^2=e^{-s/{\lambda }_f({\alpha }_0)},\label{eq:trans}
\end{equation} 
\noindent where $s$ is the local thickness of the sample. ${\lambda }_f$ depends on ${\alpha }_0$, the aperture of the objective lens, as electrons scattered to higher angles will not be detected. 

In a TEM a sample is typically located on some kind of support film or embedded in a homogeneous medium, as for example in Cryo-EM, where the medium is vitrified water. The transmission of the sample $\Ts$ and the background film or medium $\Tb$ can be calculated according to \eqref{eq:trans}. Assuming shot-noise limited electron detection, the $\SNR$ of multi-pass scattering TEM can be written as
\begin{equation}
\SNR_m=\sqrt{N_0}\frac{\left|\Ts^m-\Tb^m\right|}{\sqrt{\Ts^m+\Tb^m}}.
\end{equation}
Passing an incoming electron through a sample multiple times increases the totally applied dose a sample is exposed to and an effective multi-pass dose can be defined as 
\begin{equation}
\Deff=\frac{eN_0}{{\delta }^2}\sum^m_{i=1}{T^{i-1}_S}=\frac{eN_0}{{\delta }^2}\ \frac{1-\Ts^m}{1-\Ts}, \label{eq:Neff}
\end{equation}
which for $\Ts\to 1$ yields $D=m\frac{eN_0}{{\delta }^2}$. For $m=1$ the above equations reduce to the single-pass result. In order to identify a feature with a certain $\SNR=\SNRc$ and applying a certain effective dose $\Deff=\Deffc$, the feature size must be 
\begin{equation}
\delta_m =\frac{\SNRc\sqrt{e}}{\sqrt{\Deffc}}\frac{\sqrt{\Ts^m+\Tb^m}}{\left|\Ts^m-\Tb^m\right|}\sqrt{\ \frac{1-\Ts^m}{1-\Ts}},
\end{equation}
 which gives the multi-pass DLR. 
For highly transmissive samples ($\Ts\to 1,\ \Tb\to 1$) it scales as $1/\sqrt{m}$. Note that an image of constant resolution could be taken at an effective dose that is $m$ times lower, implying $m$ times less damage.
\subsection{$\SNR$ and DLR in Dark-Field Multi-Pass TEM.}
In dark-ground techniques the undiffracted beam is blocked and the signal now scales as $m^2\phi^2$ for a purely phase shifting specimen \cite{Born1993}. Shot-noise then scales as $m$, leading to the same scaling laws of $\SNR$ and DLR as observed in bright-field techniques.

\subsection{Multislice Simulations of Multi-Pass TEM.}

Multislice simulations were done using the methods and atomic potentials given in Kirkland \cite{Kirkland2010AtomicFactors_specialformatting}, using custom Matlab code. An ideal plane wave was propagated in alternating directions through the sample, with no wavefront aberrations applied between passes (we assume that the lenses and mirrors in the optical system can compensate for each other's aberrations). For both the protein samples and graphene, thermal smearing of \SI{0.1}{\angstrom} was applied to the atomic potentials. For graphene this was done with 32 frozen phonon configurations, while for the proteins Gaussian convolution was applied to the atomic potentials. A maximum scattering angle was enforced between each pass by applying an aperture cutoff function, equal to 20 mrad for the protein samples and 50 mrad for the graphene sample.   Inelastic losses were included by filtering out a fraction of the electron wave each pass, effectively assuming that we can filter out electrons with large inelastic losses ($>$5 eV) each pass using the optical stack. For graphene imaged at 60 kV, we assume 1.54\% inelastic loss per pass, estimated by measuring losses from an experimental STEM-EELS spectrum recorded on a NION TEM at the SuperSTEM facility. For the protein sample, we assume the inelastic losses are dominated by the vitreous ice portion of the sample.  We assumed an ice thickness of 20 nm, and a loss of roughly 5.5\% per pass at 300 kV, estimated from the literature \cite{Vulovic2013ImageMicroscopy}. The protein structures were taken from the Protein Data Bank (PDB ID: 5I4T \cite{Wagner2016CrystalSwitch.} and PDP ID: 5TOI \cite{Bruhn2016CrystalDomain.}). 
At the surface of the protein, we used the continuum model of vitreous ice given by Shang and Sigworth \cite{Shang2012}, which was implemented using 3D integration. Finally, we assumed an ideal phase plate (-$\pi$/2 phase shift of the unscattered center beam) was applied to the electron plane wave after it is coupled out of the optical cavity (a near-ideal phase plate design has been demonstrated experimentally \cite{Khoshouei2016VoltaPrx3}).

\subsection{Engineering and Design of a Multi-Pass TEM Instrument.}
While a multi-pass TEM still has to be demonstrated, the necessary components exist. Lenses and mirrors are lossless and can be used to correct for each other's aberrations \cite{Tromp2010ADesign}. Long storage times and cavity enhanced measurements have been demonstrated in charged particle traps and storage rings \cite{Peil1999,Andersen2004a}. Fast in- and out-coupling of a charged particle beam can readily be achieved using fast beam blanking or pulsed entry and exit electrodes\cite{Zajfman1997}. 
A design for a multi-pass TEM is currently under development. Proof-of-concept design simulations show that at 10 keV re-imaging to within \SI{4}{nm} is possible in a full-field all-electrostatic design using a tetrode mirror to correct for the aberrations induced by the objective.  

\end{methods}

\begin{addendum}
\item[Acknowledgements]

We thank Pieter Kruit for fruitful discussions as well as Fredrik Hage and Quentin Ramasse for providing a STEM-EELS spectrum of graphene. This research is funded by the Gordon and Betty Moore Foundation, and by work supported under the Stanford Graduate Fellowship. Work at the Molecular Foundry was supported by the Office of Science, Office of Basic Energy Sciences, of the U.S. Department of Energy under Contract No. DE-AC02-05CH11231.

\item[Author Contributions]
T.J., B.K. and M.K. conceived the technique. S.K., T.J. and C.O. performed the simulations. T.J., S.K., C.O. and M.K. analyzed the results. All authors prepared the manuscript.
\item[Competing financial interests]
The authors declare no competing financial interests.
\item[Correspondence] Correspondence and requests for materials
should be addressed to T.J.~(email: juffmann@stanford.edu).
\end{addendum}

\section*{References}

\begin{NoHyper}
    %\bibliography{Mendeley_Multipass_Microscope_Design,special_formatting}

\end{NoHyper}

\end{document}